\documentclass[11pt]{article}
\usepackage{amsmath,amssymb,amsthm}
\usepackage{graphics,epsfig}
\usepackage{hyperref}
\usepackage{natbib}
\usepackage{color}
\usepackage{graphicx}
\usepackage{caption}
\usepackage{subcaption}
\usepackage{float}
\usepackage{dsfont}
\usepackage{mathrsfs}
\usepackage{multirow}
\usepackage{comment}
\usepackage{setspace}
\usepackage{bbm}
\usepackage{bm}
\usepackage{amsfonts}
\usepackage{xcolor}
\usepackage{pslatex}
\usepackage{setspace}


\begin{document}
	
	\title{\bfseries Quantum Strategies and Economic Equilibrium*}
	
	\author{
		Arnab Kumar Maity \footnote{email:{\small\texttt{arnab.maity@pfizer.com}}}\; \footnote{Pfizer Inc., 10777 Science Center Drive, San Diego, CA 92121, USA.}}
	
	\date{\today}
	\maketitle
	
\subparagraph{Abstract.}
Abstract

\subparagraph{Key words:} .	


The quest for understanding the complex phenomena of the world has led to the development of various fields of science, each with its own methods, models, and assumptions. However, sometimes these fields can intersect and inspire each other, revealing new insights and perspectives that transcend the boundaries of their domains. In this paper, we explore one such fascinating intersection: the connection between economic theory and stochastic game dynamics, with applications to both firms and soccer players.

Economic theory is the study of how agents make decisions under scarcity and uncertainty, and how their choices affect the allocation of resources and the welfare of society. One of the central concepts in economic theory is equilibrium, which describes a state where no agent has an incentive to deviate from their chosen strategy, given the strategies of others. There are different types of equilibrium, depending on the assumptions and the context of the problem. For example, in a Walrasian framework, where prices are determined by the market, a Pareto optimal equilibrium is one where no agent can be made better off without making someone else worse off. However, in a non-cooperative setting, where agents act independently and strategically, a Nash equilibrium is one where no agent can improve their payoff by unilaterally changing their strategy.

Stochastic game dynamics is the study of how agents interact and evolve over time in a stochastic environment, where the outcomes are partly determined by chance. One of the main tools in stochastic game dynamics is the path integral, which is a mathematical technique that allows us to compute the probability of a certain outcome or event, given a set of initial and final conditions. The path integral can be seen as a generalization of the principle of least action, which states that the actual path taken by a system is the one that minimizes the action, a quantity that measures the cost or effort of the motion. The path integral sums over all possible paths, weighted by their actions, and thus captures the effects of uncertainty and fluctuations.

In this paper, we review recent papers that apply the path integral approach to two different problems in economic theory and stochastic game dynamics: the optimal control of firms and the optimal strategy of soccer players. We show how these papers use the path integral to derive novel results and insights, such as the non-cooperative feedback Nash equilibrium for firms, and the quantum formula for soccer players. We also discuss the implications and limitations of these results, and suggest some directions for future research. We hope that this paper will stimulate further interest and exploration in this exciting and interdisciplinary area of inquiry \citep{pramanik2023optimization, polansky2021motif, vikramdeo2023profiling, khan2023myb, kakkat2023cardiovascular}.

In this paper we discuss a generalized concept of cooperation by assuming that all firms behave rationally under asymmetric information. Their economic influence creates curvature in strategy spacetime. Furthermore, we assume the information set is incomplete and imperfect. By rationality, a firm uses all currently available information and resources to make its decisions. Under incomplete and imperfect information, two firms do not have any prior knowledge about each other and make guesses based on the available information in economy. Every firm has its own dynamic strategy to do business. Therefore, there exists a polygonal curved strategy space for each firm formed by the strategies taken by it historically. In this paper we include time as an important component of this curved strategy space and we define a curved strategy spacetime. 

Time is an important aspect of strategy. Decisions made at one time-point may affect the number and type of strategies at a later time. This implies that the shape of the strategy set is time dependent. Alternatively, if the market environments are stable then a firm might not need to cooperate with a new firm and the strategy may be stable. This behavior depends on the available information to the firm. If information is perfect and complete, then keeping the same strategy is sustainable. However, under asymmetric information a stable strategy is not possible and the shape of the strategy polygon is probabilistic in time.  Furthermore, we assume that the stubbornness strategy of the executive board of a firm follows a Gaussian free field with Dirichlet charge on it. Then we construct an action function based on $2$-brane and by using a Feynman type path integral approach we obtain a Wheeler-DiWitt type of equation. First order differentiation with respect to the cooperation coefficient gives us an optimal degree of geodesic cooperation. This is the main contribution of the paper in the sense that, this methodology can solve more generalized geodesic cooperative game where existing market power of a firm creates a curvature in strategy spacetime.

Early work on differential games focused on the dynamic programming method or the Hamiltonian-Jacobi-Bellman-Isaacs (HJB) method. Researchers aimed to make the concept of the value of a differential game precise and provide a rigorous derivation of the HJB equation, which often lacks a classical solution. Several studies have shown that smooth solutions do not exist for the HJB equation, and the non-smooth solutions are highly non-unique. In the 1980s, the concept of viscosity solution for the HJB equation was introduced. It characterizes the value function as the unique solution of the HJB equation with suitable boundary conditions. Viscosity solution also provides the tools to demonstrate the convergence of dynamic programming algorithms to the correct solution of the differential game and to establish the rate of convergence. Further analysis of the viscosity solution of the HJB equation in infinite dimensions is provided in various studies.

The Feynman path integral is a quantization method that uses the quantum Lagrangian function, contrasting with Schr\"odinger's quantization, which uses the Hamiltonian function. The path integral approach offers a different viewpoint from Schr\"odinger's quantization and is valuable not only in quantum physics but also in engineering, biophysics, economics, and finance. While these two methods are believed to be equivalent, their full mathematical equivalence has not been proven, mainly due to the Feynman path integral not being an integral by means of a countably additive measure. As the complexity and memory requirements of grid-based partial differential equation (PDE) solvers increase exponentially with the dimension of the system, this method becomes impractical in high-dimensional cases. An alternative approach is to use a Monte Carlo scheme, which is the foundation of path integral control. This method solves a class of stochastic control problems with a Monte Carlo method for an HJB equation, avoiding the need for a global grid of the domain of the HJB equation.

The main focus of our paper is on semi-cooperation, where $n$ identical firms exist, and each firm maximizes its dynamic profit function subject to its market dynamics. We construct a Feynman-type path integral to obtain a Schr\"odinger-type equation. By differentiating with respect to the strategy, we derive an optimal strategy. The strategy of a firm, such as its expenditure on commercials, plays a crucial role. A firm that can spend more on commercials at one time is likely to sell more products at the next time step, capturing this effect in the market share. Therefore, the strategy obtained in this framework is a feedback strategy. Cooperation is popular due to its ability to address externalities. Even the smallest cooperation considers externalities, which is at the core of bargaining power\citep{pramanik2023scoring, pramanik2020optimization, pramanik2016tail, hua2019assessing, pramanik2021effects, pramanik2023path, pramanik2022lock, pramanik2023path1, pramanik2022stochastic, pramanik2023optimal}.

We also assume that a large existing firm in an economy creates a curvature around itself by showcasing more market or political power or through advertising. Due to asymmetric information, a smaller firm has imprecise knowledge of the larger firm. Therefore, the strategy of an existing large firm should create a higher degree of curvature in the strategy spacetime to attract more fringe firms and achieve higher profit. However, if there are multiple large firms, a new fringe firm perceives multiple curvatures and breaks apart into smaller pieces in terms of its market share, leading to a scenario akin to semi-cooperation, where firm A cooperates with firms B, C, and D. When two firms have similar market power, as they come closer in the strategy spacetime, they feel the curvature in each other's strategy and form a cooperation.

There are other scenarios where curves in the strategy spacetime can be observed. For instance, during a political rally, a candidate may emphasize their political party's agenda. Supporters of the party tend to believe every word of the candidate, falling for the curvature of the strategy spacetime created by the candidate. Conversely, opponents of the candidate would not believe anything the candidate says, avoiding the curvature created by the candidate. This principle also applies to sales representatives calling customers. Salespeople often highlight the advantages of products or services, creating a curvature in the strategy spacetime of sales. If a customer is convinced of these advantages and makes a purchase, they fall for the curvature in the strategy spacetime created by the sales representative. 

This concept can be applied to sports as well. In a soccer match, each defender of a team creates a curvature around themselves such that if an opposing striker falls into it, they cannot score against that defender's team. Therefore, a striker's optimal strategy is to move beyond the "Schwarzschild radius" of each defender to score a goal.

In current times, we observe "lockdowns" of economies as a strategy to reduce the spread of COVID-19, which has claimed over 999,790 lives in the United States and more than 6 million worldwide. Many countries have implemented this strategy for all sectors of their economies except essential services such as healthcare and public safety. Different states in the United States have implemented lockdowns at different times based on their infection rates and the phase of extremely contagious transmission. Reopening has been initiated following a slowdown in the infection rate and a reduction in public activities. One negative impact of lengthy lockdowns may be a reluctance of people to engage in socioeconomic activities outside their homes due to fear of infection. This reluctance can lead to a reduction in customers for stores, potentially forcing them to reduce the number of employees or even shut down if they do not have enough inventory. Reopening a business after shutdown can be challenging, especially without sufficient government financial support.

The condition for a shutdown is determined by minimizing a healthcare cost function subject to a stochastic multi-risk Susceptible-Infectious-Recovered (SIR) model. The SIR model is a foundational model for infectious disease transmission, and almost all mathematical models of infectious disease transmission are derived from it. The deterministic part of this stochastic SIR model includes a saturated transmission rate that depends on the individual's location. Individuals in urban areas tend to have more interactions with others, leading to a higher chance of infection compared to those in rural areas. The diffusion aspect of the SIR model is relevant when individuals from rural areas visit cities and interact with others, increasing their risk of infection. Poor air quality can also contribute to respiratory illnesses, cardiovascular health issues, and reduced life expectancy. Random environmental factors such as sudden changes in air quality due to volcanic eruptions, storms, wildfires, and floods can create a more vulnerable atmosphere. Preexisting health conditions such as obesity, diabetes, hypertension, a weak immune system, and older age place individuals at a higher risk of COVID-19 infection \citep{pramanik2021optimal, pramanik2024bayes, pramanik2020motivation, pramanik2021optimization, pramanik2021consensus}

Recent research on the COVID-19 pandemic has often approached the topic from a control perspective, discussing lockdown measures and their medical, societal, and economic impacts. Various papers have explored these aspects, assuming a central planner government that can impose mandates beneficial to the population as a whole. However, it's important to note that not all individuals may comply with such mandates due to personal beliefs or circumstances. Individualistic behaviors have been studied in the context of game theory, introducing a new dimension to pandemic models.

The COVID-19 pandemic has highlighted several risk factors and comorbidities, such as obesity, diabetes, hypertension, and advanced age, which contribute significantly to its spread. Environmental factors like air pollution, temperature, and humidity also play a role. For example, increased levels of PM2.5 pollution have been associated with higher COVID-19 transmission rates. Wildfires and other events can exacerbate this issue.

In this paper, we propose a Feynman-type path integral approach to formulate a health objective function. This approach incorporates stochastic fatigue dynamics, a forward-looking stochastic multi-risk SIR model, and a Bayesian opinion network on vaccination against COVID-19. Our goal is to solve a minimization problem, $\mathbf{H}_{\theta}$, dependent on a deterministic weight, $\theta$. By using path integral control and dynamic programming tools, we can analyze and numerically solve this stochastic pandemic control model. The resulting equation resembles a Wick-rotated Schr\"odinger type equation, analogous to a Hamiltonian-Jacobi-Bellman equation and a saddle-point functional equation. This method allows for a more efficient solution compared to traditional grid-based partial differential equation solvers.

The optimal implementation of lockdown measures during the COVID-19 pandemic is a complex issue. Stricter policies can reduce infection rates but may also have significant economic and social costs. Targeted policies aimed at specific age groups or risk factors can be more effective in reducing fatalities and economic damage. The optimal timing and duration of lockdowns require careful consideration, as well as strategies to mitigate the effects of incomplete and imperfect information.

In conclusion, the use of advanced mathematical and computational methods, such as the Feynman path integral approach, can provide valuable insights into the control and mitigation of the COVID-19 pandemic. These methods offer a new perspective on pandemic modeling and can help policymakers make more informed decisions.

Public opinion plays a crucial role in shaping the response to the COVID-19 pandemic, particularly regarding vaccine mandates. In the United States, when policymakers mandated vaccination for all public sector employees, protests ensued, and many employees took leave in protest. This resistance stems from concerns about civil rights and religious beliefs. Understanding and addressing these opinions are essential for effectively managing the pandemic.

Social networks play a significant role in shaping public opinion. While there is research on social networks, limited work has explored the role of personal opinions in vaccine mandates and their influence on disease spread. Formalizing networks as simultaneous-move games, as proposed by \cite{sheng2020}, provides a framework for understanding these dynamics. This approach considers social links based on utility externalities from indirect friends and offers a computationally feasible method for analyzing large social networks.

The development of optimal strategies for firms within a competitive environment involves maximizing a stochastic Lagrangian, which represents dynamic profit functions. Unlike classical deterministic approaches, this method offers advantages in stochastic control theory. By leveraging a Feynman-type path integral approach, complex interactions among numerous firms can be effectively modeled and analyzed. This approach reduces the complexity associated with traditional Bellman and Hamiltonian methods, offering a more efficient solution for stochastic differential game problems.

Recursive methods are fundamental in dynamic economic models and other fields. They provide a framework for analyzing complex interactions and decision-making processes. However, challenges arise when dealing with forward-looking constraints, as optimal solutions may not satisfy standard principles like Pontryagin's maximum principle. Addressing these challenges requires innovative approaches, such as the promised-utility method, which transforms dynamic optimization problems into frameworks resembling HJB equations.

The Feynman path integral offers a unique quantization approach that finds applications beyond quantum physics. Its versatility and effectiveness make it a valuable tool in engineering, biophysics, economics, and finance. While mathematical equivalence with Schrödinger's quantization remains elusive, the Feynman path integral's practical advantages, such as its ability to handle high-dimensional problems, make it a promising approach for complex systems.

In conclusion, understanding public opinion, social networks, and optimal strategies is crucial for effectively managing the COVID-19 pandemic and other complex problems. Leveraging innovative approaches like the Feynman path integral can provide valuable insights and solutions in these challenging times.

The strategic analysis of soccer games involves understanding the optimal behaviors of players and teams in various situations. One key aspect is the balance between offensive and defensive strategies. For instance, after the early 1990s, Fédération Internationale de Football Association (FIFA) expressed concern over teams adopting overly defensive tactics, leading to a decrease in the total number of goals scored in matches. To address this, FIFA aimed to encourage more attacking and high-scoring matches \citep{hertweck2023clinicopathological, dasgupta2023frequent, pramanik2024estimation}.

Teams often face complex decisions regarding their strategies, influenced by factors such as the current score, time remaining in the match, and the skill level of players. One approach to analyzing these decisions is through game theory, which models the interactions between players as strategic games. For example, in a dynamic game-theoretic model, teams continuously choose between defensive and attacking formations based on the current state of the game. Research in this area has shown that teams leading the game tend to adopt more defensive strategies, especially in the second half, to protect their lead.

However, game-theoretic models in soccer face challenges due to the complexity of player behaviors and the stochastic nature of goal dynamics. Traditional models assume discrete strategies, such as attacking or defensive play, and do not consider additional dimensions of strategy, such as the choice between violent and non-violent play. Moreover, these models often rely on simplifying assumptions that may not accurately capture the nuances of player decision-making.

To address these challenges, a new game-theoretical model is proposed, utilizing the Feynman path integral method and a $\sqrt{8/3}$ Liouville Quantum Gravity (LQG) surface to represent player strategies. This approach considers players' strategies as dynamic and stochastic, evolving based on past experiences, environmental conditions, and opponent behaviors. The LQG surface provides a flexible framework for modeling these complex interactions, allowing for the analysis of optimal strategies in soccer games.

By incorporating the stochastic nature of player behaviors and goal dynamics, this model offers a more realistic representation of soccer games. It considers factors such as uncertainties due to weather, player skills, and crowd support, which can impact the outcome of a match. Through numerical simulations, the model can determine optimal strategies for players and teams, taking into account the dynamic nature of the game.

In conclusion, the strategic analysis of soccer games is a complex and dynamic process, influenced by a variety of factors. Game theory provides a valuable framework for understanding these interactions, but traditional models may need to be refined to capture the full complexity of player behaviors. The proposed model, utilizing the Feynman path integral method and LQG surface, offers a promising approach to analyzing optimal strategies in soccer games and can provide insights into the strategic decisions of players and teams.


\bibliographystyle{apalike}
\bibliography{bib}	
\end{document}